# Compact 20-pass thin-disk amplifier insensitive to thermal lensing

M. Zeyen*[a], A. Antognini[a,b], K. Kirch[a,b], A. Knecht[b], M. Marszalek[b], F. Nez[c], J. Nuber[a], R. Pohl[d], I. Schulthess[a], L. Sinkunaite[a] and K. Schuhmann[a]

[a]Institute for Particle Physics and Astrophysics, ETH, 8093 Zurich, Switzerland;
[b]Paul Scherrer Institute, 5232 Villigen PSI, Switzerland;
[c]Laboratoire Kastler Brossel, Sorbonne Université, CNRS, ENS-University PSL, Collège de France, 75005 Paris, France;
[d]Johannes Gutenberg-Universität Mainz, QUANTUM, Institut für Physik & Exzellenzcluster PRISMA, 55128 Mainz, Germany;

**ABSTRACT**

We present a multi-pass amplifier which passively compensates for distortions of the spherical phase front occurring in the active medium. The design is based on the Fourier transform propagation which makes the output beam parameters insensitive to variation of thermal lens effects in the active medium. The realized system allows for 20 reflections on the active medium and delivers a small-signal gain of 30 with $M^2 = 1.16$. Its novel geometry combining Fourier transform propagations with 4f-imaging stages as well as a compact array of adjustable mirrors allows for a layout with a footprint of 400 mm x 1000 mm.

*zeyenm@phys.ethz.ch

## 1. INTRODUCTION

High-power laser systems have become a valuable tool in many fields ranging from industrial welding[1] to gravitational wave detection[2]. Often such high-power laser systems are built in the form of a laser oscillator followed by an optical multi-pass amplifier[3,4]. Despite the continuous progress made over the last decades, most high-power multi-pass amplifiers are still limited by thermal lensing. In the case of imaging-based multi-pass amplifiers (e.g. using 4f-relay imaging from active medium to active medium) the changes in phase front curvature due to thermal lensing add up from pass to pass so that the output beam parameters strongly depend on the thermal lens at the active medium[5,6]. Compactness is also an important parameter of an optical amplifier as, generally speaking, compact systems are cheaper and less sensitive to thermal variations, seismic noise as well as vibrations caused by the cooling system.

Here we present the principle and its implementation of a novel thin-disk multi-pass architecture obtained by alternating optical segments that act as Fourier transform and optical segments that act as 4f-relay imaging. The Fourier transform segments allow a passive compensation of spherical phase front distortions occurring at the active medium[7] (Section 2). On the other hand, the presence of the 4f-segments in our multi-pass amplifier allows an increase of the number of passes while keeping the layout compact (Section 3). This hybrid system of Fourier propagations and 4f-stages allows the realization of an amplifier suited for the high-energy and high-power frontier due to the passive compensation of thermal lensing and the capability of delivering high gain in a compact layout (Section 4).

## 2. MULTI-PASS AMPLIFIER COMPENSATING THERMAL LENS EFFECTS

The advantages of a multi-pass amplifier based on the Fourier transform become most apparent when investigating the evolution of the beam size along the optical axis for small variations of the active medium dioptric power $\Delta V$ (i.e. the inverse of its focal length). For illustration purposes, the beam size evolution for two types of 2-pass amplifiers is shown in Figure 1: a) for a 4f-based amplifier, b) for a Fourier-based amplifier. The black curves show the propagation for an active medium with no focal power, whereas the red and blue curves show the beam size evolution for an active medium with positive (focusing) and negative (defocusing) dioptric power, respectively.

Conventional 4f-based multi-pass amplifiers image the beam from pass to pass on the active medium. The beam size is thus reproduced regardless of thermal lens effects (this is true for any in-coupled beam). As in most cases the same imaging optics can be used to guide multiple passes, this makes 4f-based amplifiers versatile and simple to use. However, Figure 1 also clearly shows, that the divergence of the beam is not reproduced if the active medium exhibits non-vanishing dioptric power $\Delta V$. Indeed, the phase distortions that occur at the active medium are accumulated at each pass so that the beam leaving the amplifier has a divergence that strongly depends on $\Delta V$. A known and constant (small) $\Delta V$ can in principle be partially compensated by adapting the 4f-stage[8,9]. However, especially in the case of amplifiers working in the pulsed regime, this might not be sufficient as the thermal lens of the active medium depends on running conditions (i.e. it is time dependent). Fast active compensation might be possible but complicated and expensive.

On the contrary, our proposed scheme provides passive compensation which is an easy and robust way to deal with time varying thermal lenses. The fundamental limitation of the accumulating phase front distortions at each pass on the active medium is not present in a multi-pass amplifier where the beam is Fourier transformed from active medium to active medium. This can be seen in Figure 1b). Strikingly the phase front distortions occurring in the first pass on the active medium are compensated in the second pass on the active medium, so that the out-coupled beam has very similar size and divergence as the in-coupled beam. In other words, the output beam size and divergence are independent of variations of the active medium dioptric power $\Delta V$. However, the drawback of this scheme is that it works only for specific combinations of in-coupled beam waists $w_{in}$ and focal lengths $F$ used in the Fourier transform. The relation between $w_{in}$ and $F$ can be obtained in two steps using the complex beam parameter formalism[10] where for a laser beam its complex beam parameter $q$ is defined as

$$\frac{1}{q} = \frac{1}{R} - i\frac{\lambda}{\pi w^2}, \qquad (1)$$

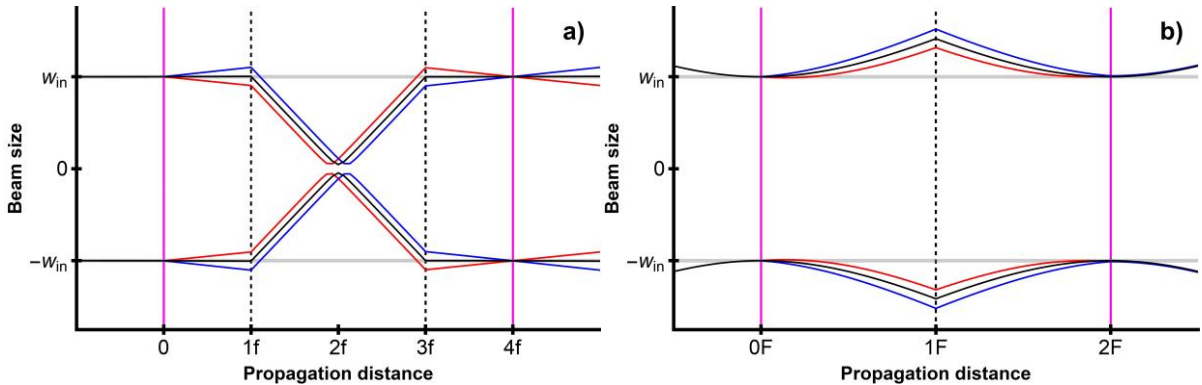

Figure 1. **a)** Scheme of the laser beam propagation in a basic 4f-setup with two passes at the active medium. The vertical magenta and dashed lines represent the location of the active medium and focusing elements, respectively. The black curves show the beam size evolution for an active medium with vanishing dioptric power. The red and blue curves show the propagation for $\Delta V = \pm\, 0.1/f$, where f is the focal length of the elements used to perform the relay-imaging. Due to the imaging properties the beam size is reproduced from active medium to active medium, but the beam divergence at the output depends strongly on $\Delta V$. **b)** Laser beam propagation in a basic Fourier transform setup with two passes at the active medium. Same color coding as for a) is used. The Fourier transform is performed by an element with focal length $F$ (e.g. a lens). In this case, for a suited choice of $F$, both beam size and divergence of the in-coupled beam are reproduced at the output independently of $\Delta V$.

with $R$ being the phase front radius, $w$ the $1/e^2$-radius of the TEM00 Gaussian beam and $\lambda$ the wavelength of the beam. The complex beam parameter $q_{out}$ at the output of an optical system can be computed knowing the 2x2 ABCD-matrix of the optical system and the input complex beam parameter $q_{in}$ using the relation

$$q_{out} = \frac{A q_{in} + B}{C q_{in} + D}, \quad (2)$$

where $A$, $B$, $C$ and $D$ are defined as the 4 elements of the ABCD-matrix characterizing the optical system. For $\Delta V = 0$, i.e., no dioptric power of the active medium, the ABCD-matrix of the Fourier-based setup shown in Figure 1b) reduces to

$$M_{FT} = M_{free} M_{lens} M_{free} = \begin{pmatrix} 0 & F \\ -\frac{1}{F} & 0 \end{pmatrix} \equiv \begin{pmatrix} A & B \\ C & D \end{pmatrix}, \quad (3)$$

where $M_{lens}$ represents the ABCD-matrix of the focusing lens needed to perform the Fourier transform and $M_{free}$ represents the ABCD-matrix of the free propagation from active medium to the lens and from the lens to the active medium, respectively. Enforcing $q_{out} = q_{in}$ and combining Equations (1), (2) and (3) the relation

$$F = \frac{\pi w_{in}^2}{\lambda} \quad (4)$$

is obtained[7]. For a given waist of the in-coupled TEM00 Gaussian beam, Equation (4) fixes the propagation length of the corresponding Fourier transform to $2F$. The beam size evolution shown in Figure 1b) fulfills this condition.

The sensitivity of a multi-pass amplifier to thermal lensing can be quantified by investigating the output beam characteristics for variations of the active medium dioptric power $\Delta V$. Figure 2a) compares the output beam size of the Fourier-based and the 4f-based amplifier of Figure 1. As noted before, 4f-imaging reproduces the beam size regardless of thermal lensing in the active medium (when aperture effects of the pumped active medium are neglected[7]), whereas the Fourier propagation allows stable replication of the input beam size at its output only for limited variations of $\Delta V$. Yet, this region of stability is practical as it makes the amplifier insensitive to typical variations of the thermal lensing. Figure 2b) compares the phase front curvature of the output beam for the Fourier-based amplifier and the 4f-based amplifier of Figure 1. For the 4f-based amplifier the phase front curvature decreases linearly with increasing $\Delta V$ as opposed to the Fourier-based propagation which exhibits a region of stability.

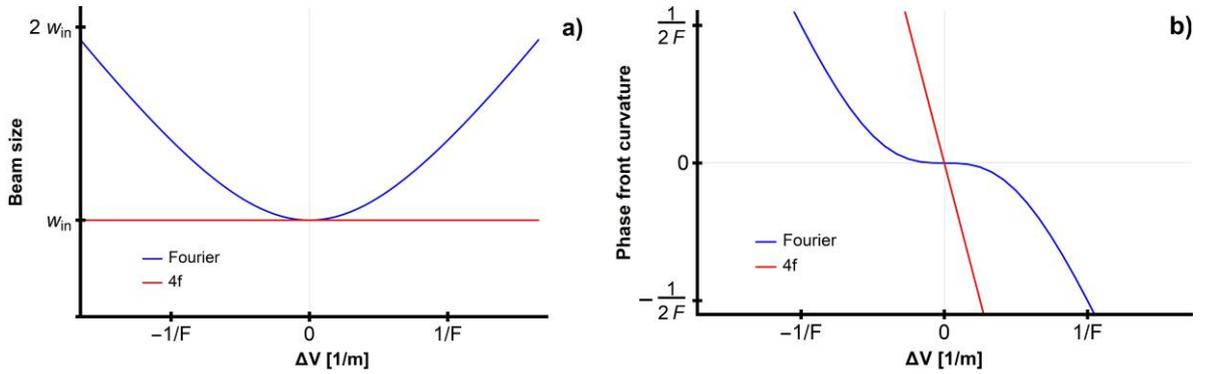

Figure 2. **a)** Beam size and **b)** phase front curvature of the output beam leaving the 2-pass amplifiers of Figure 1 for variations of the active medium dioptric power $\Delta V$.

A multi-pass amplifier stable against small variations of ΔV could in principle be realized by a succession of Fourier transform propagations and passages at the active medium as depicted in Figure 3. A practical problem arises from the fact that for high-power operation large beam sizes are needed that result in large $F$ (see Equation (4)). For example $w_{in}$ = 2.3 mm and $\lambda$ = 1030 nm lead to F ≈ 16 m so that an 8-pass amplifier would be over 230 m long. This unreasonably long propagation can be reduced by shortening the Fourier transform via back and forth propagation in a Galilean telescope[11]. An elegant multi-pass amplifier can be realized as shown in Figure 4 as a succession of

$$AM - FT - AM - SP - AM - FT - AM - SP - AM \ldots , \qquad (5)$$

where AM = active medium, SP = short propagation and FT = optical Fourier transform realized with back and forth propagation in the telescope[7]. Introduction of the SP allows for a doubling of the number of passes in a simple way, yet at the cost of a slightly increased sensitivity to variations of ΔV. This increased sensitivity can be minimized by shortening the length of the SP. A 4f-imaging instead of a short free propagation represents the optimal solution to this problem as it has zero effective propagation length. A double 4f-propagation is equally valid and it represents a practical way to implement a propagation of zero effective length (see Section 3).

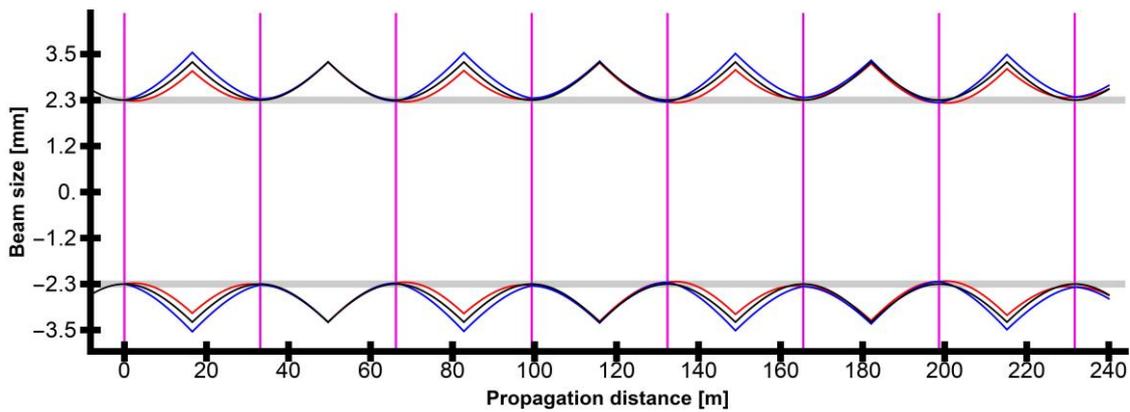

Figure 3. Beam size evolution for an 8-pass amplifier based on the Fourier transform. The purple vertical lines represent the positions of the active medium. For simplicity the focusing elements are not indicated.

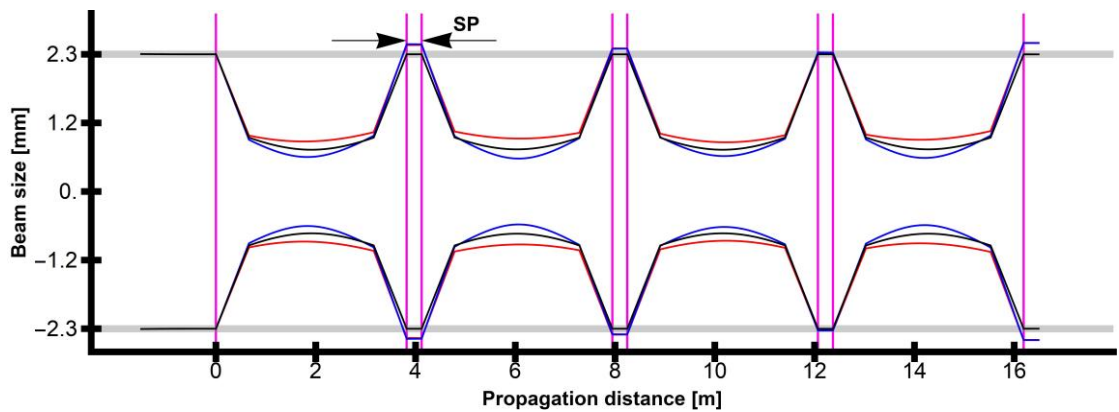

Figure 4. Beam size evolution for a Fourier-based 8-pass amplifier where the Fourier transform is achieved through a double pass in a Galilean telescope following the sequence given in Equation (5). The purple vertical lines represent the positions of the active medium. The active medium is also part of the telescope. For simplicity additional focusing elements are not indicated. Compared to Figure 3 the total propagation distance is greatly shortened.

## 3. REALIZATION OF THE HYBRID MULTI-PASS AMPLIFIER

Our active medium is a thin disk made of Yb:YAG with 7 at. % doping and a high reflective coating on its back side. The disk has a concave shape and 1 m focal length. Its thickness is 275 µm and its diameter 14 mm. The thickness of our disk is optimized for pulsed operation at low repetition rate (100 Hz) in the 100 mJ regime where amplified spontaneous emission effects have to be mitigated[12]. The pump spot diameter was 6 mm and the beam size at the disk 2.3 mm. As discussed in Section 2 the length of the Fourier transform can be shortened with the help of a Galilean telescope formed by the disk itself and a convex mirror ($f_{vex}$ = -0.5 m). The short free propagation between two Fourier transforms (see Figure 4) was implemented as a double 4f-stage having zero effective propagation length. A sketch of the beam routing through the realized multi-pass amplifier is shown in Figure 5. The 4f-stages are located on the right side of the disk, whereas the Fourier transform stages are located on its left side. The basic layout of our amplifier is thus V-shaped with the disk dividing the propagation into an arm with a short effective propagation (4f-stages) and an arm with a long effective propagation (Fourier transforms). This type of layout is also used in high power resonator designs and allows stable out-coupling of the beam from the short arm[13] (i.e. in our case after the 4f-stage).

The beam is in-coupled through a thin film polarizer (TFP) and propagates through the 4f-stage (f = 0.75 m) towards the disk where it is amplified for a first time. The beam is reflected from the disk and enters the Galilean telescope, where its size is reduced by roughly a factor of two over a distance of 0.5 m. The beam is then directed towards the mirror array and propagates a distance of 1.25 m until it reaches end-mirror M1. The λ/4-waveplate and end-mirror M1 rotates the polarization of the light by 90° and sends the beam back to a different mirror on the array. From here the beam travels back for the second time to the disk through the telescope. Any spherical phase front changes introduced by a thermal lens at the disk on the first pass are compensated at the second pass on the disk. The beam propagates back to the TFP through the 4f-stage. Since the light polarization is rotated relative to the input beam, the TFP reflects the beam towards end-mirror M2. The back-reflected beam takes a similar path as described before, except that it is rerouted by the mirror array towards the 45° mirror pair. There the polarization is not rotated so that when the beam reaches the TFP again it is reflected a second time towards end-mirror M2. All subsequent beams are routed over the 45° mirror pair and reflected at the TFP, until the second to last pass, which is directed to the λ/4-waveplate again allowing the beam to exit the amplifier at the TFP. All in all, the system has a footprint of 400 mm x 1000 mm making it quite compact.

Since the Fourier transform is not an imaging propagation, individual beams need to be handled by a mirror array (shown in Figure 5). Contrary to previous designs[14] we placed the mirror array between the Galilean telescope and end-mirror M1. This allows small incidence angles on the disk as the telescope amplifies the angular spread needed for individual beam manipulation. Additionally, the telescope reduces the beam size, allowing for a more compact design of the array.

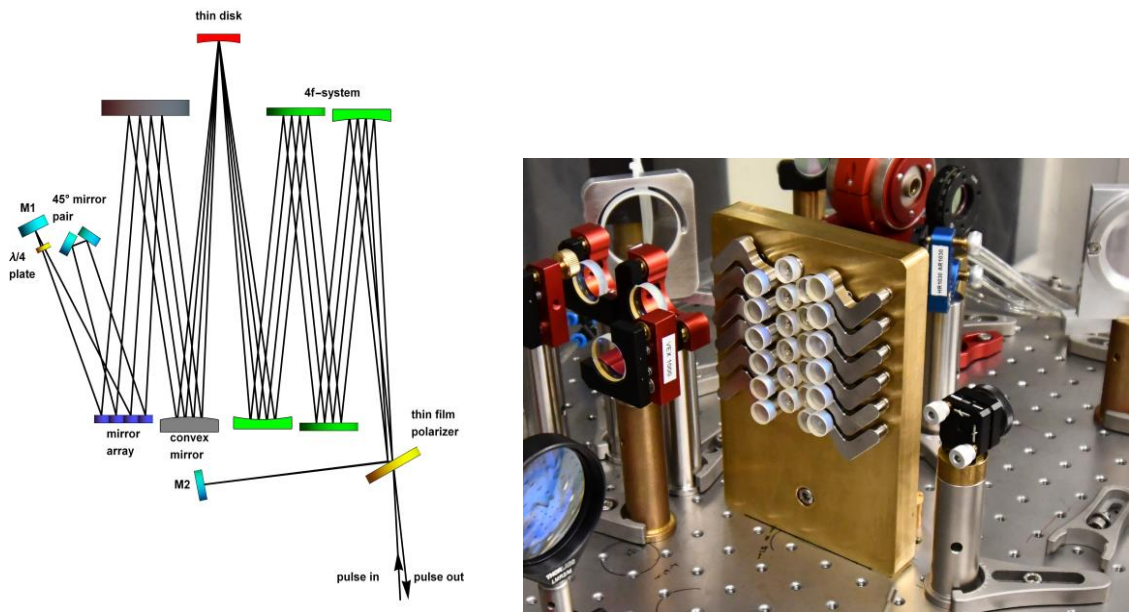

Figure 5. **Left:** Optical layout of the amplifier as viewed from above. **Right:** Picture of the compact mirror array needed for beam routing in the amplifier.

The short arm implemented as 4f-stage makes use of the small angular spread of the beams on the disk, so that all passes can be guided over the same optical elements. This allows a very compact layout of this part of the amplifier. As mentioned before the beam parameters in the short arm are insensitive to variations of ΔV which makes this part of the amplifier ideal for out-coupling. However, in our multi-pass amplifier, the only location where the beams would be sufficiently separated (not overlapping) to be extracted is at the focus of the 4f-stage. A pick-off mirror could be inserted at this position, but this choice is not suited for high power operation. For this reason we opted for the manipulation of the polarization of in- and out-coupled beams which also allows us to double the number of passes on the disk.

## 4. PRELIMINARY RESULTS AND CONCLUSIONS

At a pump power density of 3.5 W/cm$^2$ at 940 nm, the 20-pass amplifier of Figure 5 has a small signal gain of 30 at a beam quality of $M^2$ = 1.16 (see Figure 6). The interplay between Fourier segments and soft aperture effects in the pumped disk not only provides cancellation of the spherical phase changes occurring in the disk but it also acts as a mode filter[7]. This filtering leads to a good beam quality over the whole propagation through the amplifier. The measured output beam profile is presented in Figure 6b).

Summarizing, we realized a compact thin-disk multi-pass amplifier based on alternating Fourier transform propagations and double 4f-imaging propagations. Our system is not as versatile as conventional 4f-based amplifiers as it can only support the (TEM00) beam size matching the Fourier transform (see Equation (4)). However, the Fourier propagations in our amplifier yield output beam characteristics (size, divergence, profile) insensitive to variations of the thermal lens in the disk. Using Galilean telescopes as well as 4f-stages we were able to build our 20-pass amplifier on a footprint of 400 mm x 1000 mm with a beam radius on the disk of 2.3 mm. To realize this novel scheme we have developed a new compact mirror array. Opposed to other systems[9,15,16] the presented amplifier does not have a focus on any optical component thereby relaxing the problem of damage threshold for high-power operation.

The next step will be to operate the system in pulsed mode. This amplifier will be used in the laser system under development for the measurement of the ground-state hyperfine splitting in muonic hydrogen at the Paul Scherrer Institute in Villigen, Switzerland, which is the follow-up experiment of the proton radius measurement[17,18].

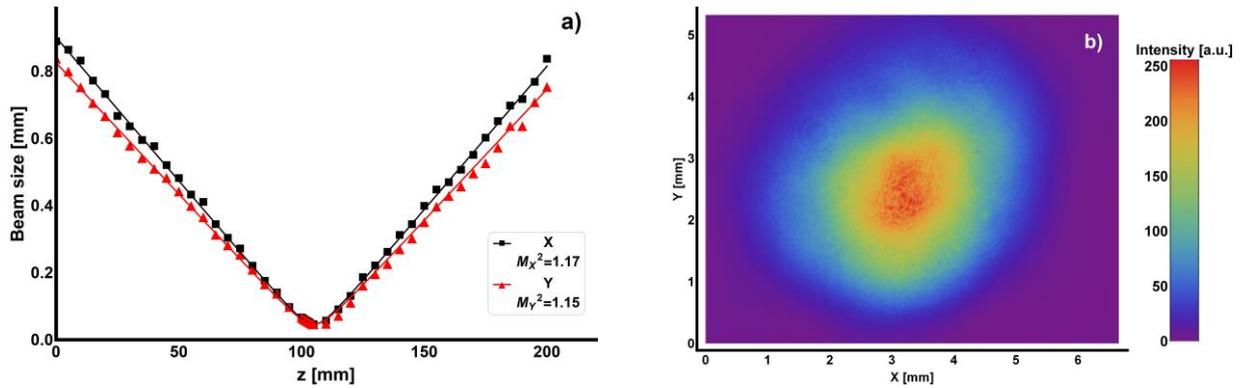

Figure 6. **a)** Measurement of the beam quality after propagation in the 20-pass amplifier of Figure 5 for a pump power density of 3.5 W/cm$^2$ and a beam radius at the disk of 2.3 mm. **b)** Intensity profile of the output beam after propagation in the 20-pass amplifier.

## ACKNOWLEDGMENTS

We acknowledge the support from the Swiss National Science Foundation Project (SNF 200021_165854) and the H2020 European Research Council (ERC) (ERC CoG. #725039, ERC StG. # 279765). The study has also been supported by the ETH Femtosecond and Attosecond Science and Technology (ETH-FAST) initiatives as part of the NCCR MUST program.